\documentclass[aps,tightenlines,nofootinbib,prd]{revtex4}
\pdfoutput=1
\usepackage{amsmath,amscd,amssymb}
\usepackage{color,epsfig}
\usepackage{xfrac}
\usepackage{latexsym}
\usepackage{cancel} 
\usepackage{mathrsfs,bigints}
\usepackage{graphicx}
\usepackage{bbm}      % double-striked fonts for sets of integer numbers etc.
\usepackage{diagbox}
\newcommand{\imu}{{\rm i}}

\newcommand{\dslash}{\partial \hskip -0.5em /}

\usepackage[dvipsnames]{xcolor}

\begin{document}

\title{Variational Approach to Excited Fermions on Kinks}

\author{Herbert Weigel$^{a)}$, Danial Saadatmand$^{a), b)}$}
\affiliation{$^{a)}$Institute for Theoretical Physics, Physics Department, 
Stellenbosch University, Matieland 7602, South Africa\\
$^{b)}$National Institute for Theoretical and Computational Sciences (NITheCS), South Africa}

\begin{abstract}
We study the back-reaction of fermion fields on the kink solution in one space and one time 
dimension. We employ a variational procedure to determine an upper limit for the minimum of the total 
energy. This energy has three contributions: the classical kink energy, the energy of valence fermions 
and the fermion vacuum polarization energy. The latter arises from the interaction of the kink with 
the Dirac sea and is required for consistency of the semi-classical expansion for the fermions. 
Earlier studies only considered the valence part and observed a substantial back-reaction. This was 
reflected by a sizable distortion of the kink profile. We find that this distortion is strongly 
mitigated when the Dirac sea is properly accounted for. As a result the back-reaction merely produces 
a slight squeeze or stretch of the kink profile.
\end{abstract}

\maketitle

\section{Introduction and Motivation}
\label{sec:intro}

Theories in one time and one space dimension ($D=1+1$) of scalar fields with degenerate vacua
often lead to static solutions that connect different vacua at the two spatial infinities. We call
them solitons (or solitary waves) when the corresponding energy density is localized. Solitons in $D=1+1$ 
models serve as role models for higher dimensional systems but can also be embedded therein. Thus they 
have numerous applications on all scales ranging from cosmic strings~\cite{Nambu:1977ag} in the electro-weak 
theory via hadron~\cite{Weigel:2008zz}, nuclear~\cite{Feist:2012ps} and condensed matter 
physics~\cite{Schollwock:2004aa,Nagasoa:2013} even to cosmology~\cite{Vilenkin:2000jqa}. A comprehensive 
summary of applications of solitons in $D=1+1$ has been compiled in the introduction of 
Ref.~\cite{MoradiMarjaneh:2022vov}.

Kink-fermion systems always have a fermion zero mode. Numerous additional fermion bound states emerge when 
the Yukawa is sufficiently large \cite{Chu:2007xh,Brihaye:2008am}. Though not kinematically stable against 
decays into free fermions, it is possible to construct local minima (or saddle points) of the static energy 
functional in which a (valence) fermion resides in an excited bound state. Not too long ago, soliton 
configurations were constructed that accounted for the back-reaction from such a higher energy valence 
level~\cite{Klimashonok:2019iya,Perapechka:2019vqv}. We reconsidered those studies and found that the 
energy of the fermion vacuum, {\it i.e.\@} the Dirac sea, which is of the same order as that of the valence 
fermion in the semi-classical expansion, contributes largely to the total energy \cite{Saadatmand:2022htx}. 
Based on that study we now attempt to identify soliton like minima of the energy functional with an excited 
valence fermion when all contributions to the fermion energy that are leading order in the semi-classical 
expansion are included. This extension is also important because we know from the bosonized Nambu-Jona--Lasino
model that, while coupling to a valence quark strongly binds the chiral soliton, the Dirac sea has a 
destabilizing effect in the sense that the energy of the polarized sea significantly increases the 
total energy \cite{Alkofer:1994ph}. However, that model does not fall into the class of renormalizable 
theories that we explore here. 

Some time ago self-consistent configurations from the binding to a single fermion bound state omitting 
the Dirac sea were considered in chiral quark models \cite{Friedberg:1976eg,Kahana:1984dx,Jain:1988ix}, 
by coupling the bound state to a magnetic monopole \cite{Callan:1982au} as well as in variants of
the electro-weak theory \cite{Nolte:1993jt}; though only for the lowest energy bound state. For 
models with fermion couplings the necessity of including the Dirac sea was later pointed out for 
one~\cite{Farhi:2000ws} and three space dimensions \cite{Farhi:2001kh}. However, cases in which
the coupling goes to an excited level are still interesting and may cause major deformations
of the kink even when the Dirac sea is included. This is a major objective of the present study. 
We note that indeed excited fermion levels play their roles in physics. For example, in the 
MIT bag model \cite{Chodos:1974je} the Roper (1440) resonance is associated with a radially 
excited quark level \cite{Bowler:1977aj,Guichon:1985ny}.

This short report is organized as follows. In Sect.\@ \ref{sec:model} we introduce the model
and discuss the classical energy of the kink as well the coupling between the kink and a single
fermion mode. In Sect.\@ \ref{sec:EF} we examine the fermion contribution to the energy with
emphasis on the Dirac sea contribution in the context of the semi-classical expansion. 
Our numerical results are contained and discussed in Sect.\@ \ref{sec:NR} while we briefly
conclude in Sect.\@ \ref{sec:Con}.

\section{The Model}
\label{sec:model}

In $D=1+1$ the scalar field $\Phi$ is dimensionless and the fermion spinors $\Psi$ have
canonical energy dimension $\frac{1}{2}$. We take the Yukawa coupling constant $g$ to be  
dimensionless and write the Lagrangian as
\begin{equation} 
\mathcal{L}=\frac{1}{2}\partial_\mu\Phi\partial^\mu\Phi
-\frac{\lambda}{4}\left(\Phi^2-\frac{M^2}{2\lambda}\right)^2
+\imu \overline{\Psi}\dslash\Psi-g\sqrt{\frac{\lambda}{2}}\,\overline{\Psi}\Phi\Psi\,. 
\label{eq:lag}\end{equation}
The scalar (or Higgs) coupling constant $\lambda$ has dimension energy squared and $m=\frac{gM}{2}$ 
is the fermion mass which arises from spontaneous symmetry breaking that generates the vacuum 
expectation value $\langle \Phi \rangle=\frac{\pm M}{\sqrt{2\lambda}}$. Scalar fluctuations about 
$\langle \Phi \rangle$ have mass $M$.

In order to find the most generic, {\it i.e.} parameter independent, formulation and
also for numerical practicality it is appropriate to introduce dimensionless quantities:
\begin{equation}
\Phi(t,x)=\frac{M}{\sqrt{2\lambda}}\,\phi(\tau,\xi)
\quad {\rm and}\quad 
\Psi(t,x)=\sqrt\frac{M}{2}\,\psi(\tau,\xi)\,,
\quad {\rm where}\quad 
(\tau,\xi)=\frac{M}{2}(t,x)\,.
\label{eq:scale1}\end{equation}
We have introduced the factor $\frac{1}{2}$ in the dimensionless coordinate so that the kink,
$\phi_{\rm K}(\xi)=\tanh(\xi)$, is the soliton solution to the field equation for $\phi$ when $g=0$.
Choosing $\gamma^0=\sigma_1$ and $\gamma^1=\imu\sigma_3$ as the representation for the Dirac matrices,
the stationary Dirac equation then is an eigenvalue equation for the dimensionless energy\footnote{In
physical variables the separation leading to the stationary equation is $\Psi(t,x)={\rm e}^{-\imu Et}\Psi(x)$.}
$\epsilon=\frac{2E}{M}$
\begin{equation}
\epsilon \psi(\xi)=h\psi(\xi)\qquad {\rm with} \qquad
h=-\imu\sigma_2\partial_\xi+g\phi(\xi)\sigma_1\,.
\label{eq:scale2}\end{equation}
The normalization condition is $\bigintsss d\xi\,\psi^\dagger(\xi)\psi(\xi)=1$. Fortunately, in our 
approach we will not need to construct these spinors; but only the eigenvalues $\epsilon$ which
can be discrete and, above threshold $g$, continuous. We try to keep the notation simple and
write the labels for these energies (and the corresponding eigen-spinors) only when necessary. In terms 
of upper ($u$) and lower ($v$) spinor components in $\psi=\begin{pmatrix}u\\[-1mm]v\end{pmatrix}$,
Eq.~(\ref{eq:scale2}) reads (primes denote derivatives with respect to $\xi$) 
\begin{equation}
u^\prime=\epsilon v -g\phi u \qquad {\rm and}\qquad
v^\prime=-\epsilon u +g\phi v\,. 
\label{eq:scale3}\end{equation}
The fermion quantum effects on $\phi(\xi)$ are non-local and when they are implemented the field equation 
 $\phi(\xi)$ is not a (simple) differential equation. However, for a given profile we have the classical energy
\begin{equation}
E_{\rm cl}=\frac{M^3}{4\lambda}\epsilon_{\rm cl}\qquad{\rm with}\qquad
\epsilon_{\rm cl}=
\int_0^\infty d\xi\,\left[\phi^{\prime2}(\xi)+\left(\phi^2(\xi)-1\right)^2\right]\,.
\label{eq:scale4}\end{equation}
Any legitimate soliton profile connects the vacuum expectations values $\langle\phi\rangle=\pm1$
between negative and positive spatial infinity and is anti-symmetric under spatial 
reflection. Then the solutions to Eq.~(\ref{eq:scale3}) separate into two channels:
the one with positive intrinsic parity has even $u$ and odd $v$, while the negative
intrinsic parity channel has it the other way round. As in Ref.~\cite{Klimashonok:2019iya}
we refer to these channels as $A$-and $B$-type solutions or configurations. Additional 
integer labels on $A$ and $B$ count the number of zero-crossings of $u$ on the half-line 
$x\ge0$, including the one at $x=0$ for the $B$-type solution. In this notation the zero 
mode of the kink is an $A_0$ solution.

\section{The Fermion Energy Functional}
\label{sec:EF}

Here we detail the treatment of the fermion contribution to the energy starting from the 
effective action for fermions interacting with a static background potential. In our case
that potential is generated by the soliton $\phi(\xi)$.

\subsection{Formal Considerations}
\label{sec:FCon}

Fermions that interact with a static background are subject to a Dirac equation of
the form $\left(\imu\partial_t-h\right)\Psi=0$ and their effective action formally, 
{\it i.e.\@} ignoring the important regularization, is \cite{Reinhardt:1989st,Alkofer:1994ph}
\begin{equation}
\mathcal{A}=\frac{T}{2}\sum_\nu|E_\nu|
-\imu\ln \sum_{\{\eta_\nu\}}{\rm exp}\left[-\imu T\sum_{\nu}\eta_\nu|E_\nu|\right]\,,
\label{eq:EF1}
\end{equation}
where the $E_\nu$ are again the eigenvalues of the Dirac Hamiltonian $h$ and $T$ is an
arbitrarily large time interval discretizing the eigenvalues of $\imu\partial_t$. The outer 
sum in the second term runs over all possible sets of occupation numbers $\eta_\nu=\pm1$ for
the single fermion levels. When singling out a particular set of occupation numbers, say
$\{\overline{\eta}_\nu\}$ this outer sum is omitted and we extract the (unregularized) fermion 
energy functional
\begin{equation}
E_F(\overline{\eta}_\nu)=-\frac{1}{2}\sum_\nu|E_\nu|+\overline{\eta}_\nu|E_\nu|\,.
\label{eq:EF2}
\end{equation}
The two contributions on the right hand side are the vacuum and valence energies, respectively.
Upon comparison with the free case without a static background the vacuum energy turns into the
vacuum polarization energy that we will regularize and renormalize utilizing spectral 
methods \cite{Graham:2009zz} in Subsect.\@ \ref{sec:FVPE}. We furthermore note that the conserved 
fermion number is
\begin{equation}
N_F(\overline{\eta}_\nu)=\sum_\nu\left(\overline{\eta}_\nu-\frac{1}{2}\right){\rm sign}(E_\nu)\,.
\label{eq:EF3}
\end{equation}
For a prescribed fermion number therefore the global energy minium corresponds to a specific 
set of occupation numbers. Unless the vacuum energy of a self-consistent configuration varies 
strongly with the selection of occupation numbers, this specific set fills levels starting from 
the most strongly bound one. Recently, however local energy minima (or at least saddle points) 
have been discussed for which the non-zero occupation numbers concern the first, second or third 
excited single particle levels \cite{Klimashonok:2019iya,Perapechka:2019vqv,Saadatmand:2022htx}. 
This scenario is the central objective of our project.

These formal considerations clearly show that fermion valance and vacuum energies must be 
treated on equal level. Yet, Eq.~(\ref{eq:EF2}) only contains the fermion one-loop contribution to 
the energy. The loop-counting parameter is the inverse of the ratio of the scales for the classical 
energy and the fermion energy eigenvalues, {\it i.e.\@} $\frac{M^2}{\lambda}$. Hence for this 
approach to produce reliable results we focus on $M^2>\lambda$. Furthermore we will omit 
quantum corrections originating from the scalar field. That is, we assume that the fermion
quantum corrections dominate the scalar ones. This is reliable when the fermion energy eigenvalues
are strongly skewed by the background potential, which happens to be the case when the Yukawa 
coupling is large: $g\gg1$.

\subsection{Bound States}
\label{sec:Bstates}

The Dirac equation has discrete, normalizable solutions with $|\epsilon|<g$. The stronger the Yukawa 
coupling, the more of these solutions exist \cite{Chu:2007xh,Brihaye:2008am}. When $\phi(\xi)$ is odd 
under spatial reflection they have definite parity. We already mentioned that in the notation of 
Refs.~\cite{Klimashonok:2019iya,Perapechka:2019vqv} these parity channels are called $A$ and $B$-type. 
An additional, integer label on the capital letters counts the bound states in a given parity channel. 
Then $A_0$ is the configuration with the most strongly bound fermion mode occupied. This mode has energy 
eigenvalue zero and is always present, no matter what the Yukawa coupling is. The second most bound 
fermion mode has opposite parity and its explicit occupation defines the $B_1$ configuration, followed 
by a bound state with the same parity as the zero mode. Its occupation defines the $A_1$ configuration, etc.\@ 
\cite{Klimashonok:2019iya,Perapechka:2019vqv,Saadatmand:2022htx}. We will maintain that notation 
for the particular choices of $\overline{\eta}_\nu$.

In the numerical simulation we find the energy eigenvalues by integrating the differential equations~(\ref{eq:scale3}) 
from the origin with initial conditions suitable for either the $A$- or $B$- configurations to some intermediate 
coordinate, $\xi_m$. Furthermore we integrate from a large distance, say $\xi_{\rm max}\gg\xi_m$ with initial
conditions for exponentially decaying spinor components $u$ and $v$ to $\xi_m$ as well. Only for certain 
parameters $\epsilon$ in Eq.\@ (\ref{eq:scale3}) it is possible to match these solutions at $\xi_m$. These 
$\epsilon$ values are the searched-for energy eigenvalues. We verify that the resulting bound state energies are 
not sensitive to the choice of $\xi_m$ ($\xi_{\rm max}$) when it is taken not too large (small).

\subsection{Fermion VPE}
\label{sec:FVPE}
In this section we renormalize the divergent vacuum part of the fermion energy in Eq.~(\ref{eq:EF2}) 
and express it in terms of scattering data. First we measure this energy relative to the 
$\Phi\equiv\langle \Phi\rangle$ case so that it turns into the vacuum polarization energy (VPE)
\begin{equation}
E_{\rm VPE}=-\frac{1}{2}\sum_\nu\left(|E_\nu|-|E^{(0)}_\nu|\right)\Big|_{\rm ren.}\,.
\label{eq:VPE1}
\end{equation}
The sum contains bound and scattering states. The latter is expressed as a momentum integral over 
continuum energies weighted by the change in the density of states generated by $\Phi\ne\langle\Phi\rangle$. 
That change is computed from scattering data according to the Friedel-Krein formalism \cite{Faulkner:1977zz}. 
The main ingredient is the momentum derivative of the phase shift that describes the fermion scattering about 
the background potential. Renormalization (indicated by the subscript) is then 
accomplished by subtracting sufficiently many terms the Born series, which is an expansion in the strength 
of the potential, from the integrand and adding those pieces back in as Feynman diagrams, which also combine 
to an expansion in that strength. Those Feynman diagrams are combined with the counterterms of the chosen 
renormalization scheme. The counterterm Lagrangian may only contain terms that arise from varying parameters
(or scaling fields) in Eq.\@ (\ref{eq:lag}) and the coefficients of these terms may not depend on the 
peculiarities of the fields.

In the present application we still have to account for the fact that with $\phi(\pm\infty)=\pm1$ the fermion 
mass terms, as induced by spontaneous symmetry breaking, have opposite signs at positive and negative spatial
infinity. To this end we note that for a static system which is invariant under charge conjugation we can formally
write the effective action, from which the VPE is extracted, as
\begin{equation}
\int \frac{d\omega}{2\pi}\,{\rm Tr}{\rm Log}\left[\omega-h\right]
=\frac{1}{2}\int \frac{d\omega}{2\pi}\,{\rm Tr}{\rm Log}\left[\omega^2-h^2\right]\,.
\label{eq:Aeff}\end{equation}
Thus the VPE of this fermion system can be obtained from the average VPE of two scalar 
systems~\cite{Graham:1999pp} associated with Eq.~(\ref{eq:scale3}) since
\begin{equation}
h^2=\begin{pmatrix} -\partial^2_\xi -g\phi^\prime+g^2\phi^2 & 0 \cr 
0 & -\partial^2_\xi +g\phi^\prime+g^2\phi^2 \end{pmatrix}\,.
\label{eq:vpe1}\end{equation}
The two potentials are straightforwardly read off as
\begin{equation}
V_S=g^2\left(\phi^2-1\right)-g\phi^\prime
\qquad \mbox{and}\qquad
\widetilde{V}_S=g^2\left(\phi^2-1\right)+g\phi^\prime\,.
\label{eq:vpe2}\end{equation}
They are invariant under spatial reflection. For any scalar potential $\sigma$ in $D=1+1$ 
dimensions with that invariance the renormalized VPE is computed as
\begin{equation}
\epsilon_{\rm VPE}[\sigma]=\int_{0}^\infty \frac{d\tau}{2\pi}\,\left\{
{\rm ln}\left[G(t,0)\left(G(t,0)
-\frac{1}{t}G^\prime(t,0)\right)
\right]-\frac{\langle \sigma\rangle}{t}\right\}_{t=\sqrt{\tau^2+g^2}}\,.
\label{eq:vpe4} \end{equation}
Let us outline the connection of Eq.\@ (\ref{eq:vpe4}) with scattering data:
For real momenta $k$ the scattering phase shift is the phase of the Jost function, $f(k,\xi)$.
The Jost function is the solution of the wave-equation with potential $\sigma(\xi)$ that 
obeys the boundary condition $\lim_{\xi\to\infty}f(k,\xi)\,{\rm e}^{-\imu k\xi}=1$. 
Factorizing the plane wave part $f(k,\xi)={\rm e}^{\imu k\xi}\overline{G}{(k,\xi)}$ defines 
$G(t,\xi)=\overline{G}{(\imu k,\xi)}$ by analytic continuation. It solves the ordinary 
differential equation
\begin{equation}
G^{\prime\prime}(t,\xi)=2tG^\prime(t,\xi)+\sigma(\xi)G(t,\xi)\,,
\label{eq:vpe3}\end{equation}
subject to the boundary condition $\lim_{\xi\to\infty}G(t,\xi)=1$. Eq.\@ (\ref{eq:vpe4})
applies to systems with $\sigma(-\xi)=\sigma(\xi)$ and thus decouples into parity channels.
The first factor under the logarithm stems from odd parity (wave-function vanishes at the origin)
while the second one originates from even parity (derivative of the wave-function vanishes 
at the origin).

As few further remarks are 
in order to explain Eq.~(\ref{eq:vpe4}). First, the last term under the integral proportional to
$\langle \sigma\rangle=\bigintsss_0^\infty d\xi\, \sigma(\xi)$ is the Born approximation to the
logarithm. The corresponding Feynman diagram is fully canceled by a counterterm within the 
no-tadpole renormalization scheme. Second, the physical momentum has been analytically continued 
into the upper half complex plane. Evaluating the momentum integral as a contour integral has 
contributions from the branch cut along $k=\imu t$ in the dispersion relation $\epsilon=\sqrt{k^2+g^2}$ with 
$t>g$ and the poles arising from the logarithmic derivative of the Jost function since the Jost function
vanishes at the complex momenta of the bound state energies. Then the third important feature is that 
the latter contributions exactly cancel the bound state part in the sum of Eq.~(\ref{eq:VPE1}).
More details are given in the reviews \cite{Graham:2009zz,Graham:2022rqk}.

Restoring physical dimensions finally yields
\begin{equation}
E_{\rm VPE}=-\frac{M}{4}\left(\epsilon_{\rm VPE}\left[V_S\right]
+\epsilon_{\rm VPE}\left[\widetilde{V}_S\right]\right)\,.
\label{eq:vpe5} \end{equation}

We also observe that the Born subtraction in Eq.\@ (\ref{eq:vpe4}) involves the integral
$$
\langle V_S +\widetilde{V}_S\rangle=2g^2\int_0^\infty d\xi\,\left(\phi^2-1\right)\,.
$$
It also arises from a counterterm which compensates changes of the vacuum expectation value 
$\langle\Phi\rangle^2=\frac{M^2}{2\lambda}$ in Eq.~(\ref{eq:lag}). Hence Eq.~(\ref{eq:vpe5}) 
indeed implements the no-tadpole renormalization condition which requires that the (fermion) quantum 
corrections do not alter this expectation value. 

\section{Numerical Results}
\label{sec:NR}

The model parameters are the mass $M$, the Higgs coupling $\lambda$ and the Yukawa coupling 
constant $g$. With the scaling in Eq.~(\ref{eq:scale1}) we factor out an overall constant $M$
from the energy so that the relevant model parameters are $g$ and the dimensionless ratio 
of the energy scales: $\alpha=\frac{M^2}{2\lambda}$. This ratio weighs the classical vs.\@ vacuum 
polarization energies and its inverse plays the role of a loop counting parameter. For the 
numerical analysis it obviously suffices to choose, {\it e.\@ g.\@} $M=2$ and scan the $g$-$\lambda$ 
parameter space; or equivalently the $g$-$\alpha$ space. This becomes obvious from the expression 
for the total energy
\begin{equation}
E_{\rm tot}[\Phi]=E_{\rm cl}+\sum_\nu \overline{\eta}_\nu |E_\nu| +E_{\rm VPE}
=\frac{M}{2}\left\{\alpha\epsilon_{\rm cl}+\sum_\nu \overline{\eta}_\nu|\epsilon_\nu|
-\frac{1}{2}\left(\epsilon_{\rm VPE}\left[V_S\right]
+\epsilon_{\rm VPE}\left[\widetilde{V}_S\right]\right)\right\}\,,
\label{eq:Etot}\end{equation}
as $g$ and $\alpha$ are the only the model parameters entering the factor in curly brackets.
An important observation is that the classical and the fermion energies scale differently with 
the model parameters. Hence the choice of the particular relation $M=\sqrt{2\lambda}$, that 
was assumed in Ref.\@ \cite{Klimashonok:2019iya}, may obscure important information \cite{Saadatmand:2022htx}.

For the particular parameters $M=\lambda=g=2$ our model matches the super-symmetric one of
Ref.\@ \cite{Graham:1999pp}. For this case our numerical simulation yields $E_{\rm VPE}=0.3479$
when substituting $\phi=\phi_{\rm K}$ in $V_S$ and $\widetilde{V}_S$. This agrees well with the
analytic result $\frac{M}{\pi}\left(1-\frac{\pi}{4\sqrt{3}}\right)$ found in that 
super-symmetric model and confirms the validity of our simulation\footnote{The computations
of Ref.\@ \cite{Graham:1999pp} make ample use of scattering data; so do we. Additionally we 
use the analytic properties of the Jost function to write the VPE as a single integral
over imaginary momenta. Note also that Ref.\@ \cite{Graham:1999pp} computes the VPE 
for two widely separated kinks and therefore has an additional factor two in Eq.\@ (21).}.

The total energy is solely a functional of the scalar field $\Phi$ (or $\phi$ in dimensionless 
variables) since for any given scalar profile the fermion contributions are determined by the Dirac 
equation~(\ref{eq:scale3}) and/or its downstream scattering equation~(\ref{eq:vpe3}). As already 
mentioned, this extremal condition cannot be formulated as a set of differential equations due to 
the non-local structure of the VPE. In a non-renormalizable model with a finite ultra-violet cut-off,
$\Lambda$, a finite and countable set of eigen-functions of the stationary Dirac equation exists 
and the functional derivative $\frac{\delta \mathcal{A}}{\delta \phi(\xi)}$ can be computed via
the Feynam-Hellmann theorem. Eventually this produces an implicit field equation that can be
solved self-consistently. Adopting that method to compute the VPE requires the limit 
$\Lambda\,\to\,\infty$ on top of the self-consistent approach. Not only does that seem numerically
infeasible\footnote{There have been attempts to pursue that program \cite{Diakonov:1995xz}, 
however, the relation between the gauge invariant proper-time regularization and the sharp
cut-off for $\mathcal{A}$ is unclear.}
it also spoils the nice features of the spectral method, Eq.\@ (\ref{eq:vpe4}), that
all entries are ultra-violet finite and the renormalization conditions can be unambiguously 
implemented. Having discussed that, we therefore consider a parameterization for the scalar profile
that is modeled after the kink $\phi_{\rm K}=\tanh(\xi)$ with a rational function as 
correction\footnote{We write absolute values for the odd powers to maintain the reflection property
$\phi(-\xi)=-\phi(\xi)$. The actual calculation is performed on the half-line $\xi\ge0$ and
the absolute value sign can be ignored.}
\begin{equation}
\phi(\xi)=\frac{\xi^2+a|\xi|+b}{\xi^2+d|\xi|+e}\,\tanh(c\xi)\,.
\label{eq:var1}\end{equation}
For notational convenience we define the set of parameters as
\begin{equation}
\mathbb{P}=\left\{a,b,c,d,e\right\}\,.
\label{eq:defP}\end{equation}
The rational function in Eq.\@ (\ref{eq:var1}) is a Pad\'e approximation to the deviations from 
the purely scalar kink subject to the condition that it approaches unity at spatial infinity. Pad\'e 
approximations converge quickly \cite{Jr:2012} so that only few parameters are needed for an effective 
variational approach. Indeed, we will observe that the above {\it ansatz} reproduces the strongly 
distorted kinks from Refs.\@ \cite{Klimashonok:2019iya,Perapechka:2019vqv,Saadatmand:2022htx} very 
well. In addition to the rational function we introduce the variational parameter $c$ for the extension 
of the profile function.

For a given set of occupation numbers $\{\overline{\eta}_\nu\}$ we then compute $E_{\rm tot}$ for numerous 
values of the five variational parameters in $\mathbb{P}$ and identify the minimal value. To do 
so, we start with a profile function that is either close to the kink or close to one of the 
solutions constructed in Refs.\@ \cite{Klimashonok:2019iya,Perapechka:2019vqv,Saadatmand:2022htx}.
We then apply a simple steepest descent algorithm to that first choice. Eventually this will converge
to a minimum which then will be an upper bound to the actual minimum of $E_{\rm tot}$ because 
the variational space is limited. 

To gauge the quality of this fitting function we will reconsider the minimization of the 
reduced energy functional
\begin{equation} 
E_{\rm red}=E_{\rm cl}+E_{\rm val}=E_{\rm cl}+\frac{M}{2}\sum_\nu \overline{\eta}_\nu\epsilon_\nu\,.
\label{eq:Ered}\end{equation}
That minimum was previously \cite{Klimashonok:2019iya,Perapechka:2019vqv,Saadatmand:2022htx} constructed 
by self-consistently solving the Dirac equation~(\ref{eq:scale3}) together with the differential equation
\begin{equation}
\phi^{\prime\prime}=2\phi\left(\phi^2-1\right)+\frac{4g\lambda}{M^2}\,{\rm sign}(\epsilon)uv\,,
\label{eq:DEQred}\end{equation}
in which $u$ and $v$ are the spinor components of the level for which $\overline{\eta}_\nu=1$ normalized 
to $\bigintsss d\xi\left(u^2+v^2\right)=1$. Here self-consistent refers to the condition that the profile 
function in the Dirac equation with that particular energy eigenvalue is also the solution to 
Eq.\@ (\ref{eq:DEQred}).  Those self-consistent profiles turned out to 
significantly deviate from the kink. Since the VPE typically mitigates the binding from the 
occupied levels, we expect the actual solution to lie between the kink and those strongly distorted
kink profiles. Hence reproducing the latter by the above fitting function to a high precision
will justify the parameterization in Eq.~(\ref{eq:var1}). In a first step we therefore consider 
the case $M=\lambda=2$ for which we constructed self-consistent solutions that minimize the reduced 
energy functional $E_{\rm red}$ earlier~\cite{Saadatmand:2022htx}. Subsequently the VPE is computed for 
this construction. The numerical results for the $B_1$ ($\overline{\eta}_\nu=1$ for the first excited 
level with negative parity, all other $\overline{\eta}_\nu=0$) and $A_1$ ($\overline{\eta}_\nu=1$ for 
the first excited level with positive parity, all other $\overline{\eta}_\nu=0$) configurations 
are listed in table \ref{tab1} for the choice $g=4$.
\begin{table}[tb]
\begin{tabular}{c|c|c|c|c}
$B_1$ &~~~~$E_{\rm cl}$~~~~&~~~~$E_{\rm val}$~~~~&~~~~$E_{\rm VPE}$~~~~&$~~~E_{\rm tot}$~~~\\ 
\hline
 & 2.223 & 1.041 & 2.029 & 5.300\\
&\multicolumn{4}{|c}{self-consistent minimum of $E_{\rm red}$}\\
\hline
& 2.277 & 1.041 & 2.043 & 5.361\\
&\multicolumn{4}{|c}{$\mathbb{P}_1=\left\{-0.758,-0.063,0.903,-1.022,1.018\right\}$}\\
\hline
& 2.171 & 1.134 & 1.987 & 5.293\\
&\multicolumn{4}{|c}{$\mathbb{P}_2=\left\{-0.733,-0.013,0.917,-0.924,0.964\right\}$}\\
\end{tabular}\hspace{1cm}
\begin{tabular}{c|c|c|c|c}
$A_1$ &~~~$E_{\rm cl}$~~~&~~~$E_{\rm val}$~~~&~~~$E_{\rm VPE}$~~~&$~~~E_{\rm tot}$~~~\\
\hline
 & 2.079 & 2.071 & 2.000 & 6.150\\
&\multicolumn{4}{|c}{self-consistent minimum of $E_{\rm red}$}\\
\hline
& 2.095 & 2.065 & 2.001 & 6.163\\
&\multicolumn{4}{|c}{$\mathbb{P}_1=\left\{-2.163,1.395,0.759,-2.304,1.988\right\}$}\\
\hline
& 2.067 & 2.091 & 1.983 & 6.142\\
&\multicolumn{4}{|c}{$\mathbb{P}_2=\left\{-2.151,1.402,0.760,-2.263,1.958\right\}$}
\end{tabular}
\caption{\label{tab1}Comparison of energies from the fitting function and the self-consistent solution to the 
reduced problem, Eqs.~(\ref{eq:scale3}), (\ref{eq:Ered}) and~(\ref{eq:DEQred}) for the $B_1$ and $A_1$ configurations 
using $g=4.0$ and $M=\lambda=2$: $\mathbb{P}_1$ is the fit to the self-consistent solution, $\mathbb{P}_2$ is the 
variational solution. The column labeled $E_{\rm val}$ denotes $\frac{M}{2}\sum_\nu \overline{\eta}_\nu\epsilon_\nu$ 
for the respective configurations.}
\end{table}
Obviously the fit reproduces the results from the self-consistent approach convincingly well, in 
particular for the fermion ingredients. As a matter of fact, we have considered two scenarios, the first, 
labeled $\mathbb{P}_1$, is a parameter fit to the self-consistent solution and the second, $\mathbb{P}_2$,
is the variational minimum to $E_{\rm red}$. Either variational profile essentially equals the self-consistent 
one.  This is the case for both the $B_1$ and $A_1$ configurations. Surprisingly those profiles exceed 
$\phi=1$ at some moderate distance and approach the asymptotic vacuum expectation value from above
as $\xi\to\infty$. We conclude that the fitting function, Eq.~(\ref{eq:var1}) is indeed a well-suited 
variational {\it ansatz} to approximate the scalar profile which minimizes the total energy, Eq.~(\ref{eq:Etot}).

There is a subtlety with this parameterization, though. Asymptotically the 
profile behaves as $\phi(\xi)\sim1+\frac{a-d}{|\xi|}$ which causes the integral in the Born approximation
$$
2g^2\int_0^{\xi_{\rm max}} d\xi\,\left[\phi^2(\xi)-1\right]
$$
to logarithmically diverge as $\xi_{\rm max}\to\infty$. The Born approximation has been introduced to 
cancel the large $t$ component of the logarithm in Eq.~(\ref{eq:vpe4}). Since higher order terms of the 
Born series are finite when $\xi_{\rm max}\to\infty$, we get a sensible result for Eq.\@ (\ref{eq:Etot}) 
when we integrate the differential equation (\ref{eq:vpe3}) between zero and the very same $\xi_{\rm max}$.
We have confirmed that once $\xi_{\rm max}$ is taken large enough, the numerical simulation
of Eq.~(\ref{eq:vpe4}) is stable against further variations of $\xi_{\rm max}$. In the no-tadpole
scheme that first order contribution is exactly removed and there is no further problem. Although
the chosen parameterization may not be ideal asymptotically, the results from table \ref{tab1}
corroborate that it is nevertheless suitable. There is no such problem for the bound states whose
wave-functions decay exponentially\footnote{This is similar to the $S$-wave bound states for the 
Coulomb problem in the Schr\"odinger equation.}.

\begin{figure}[t]
\centerline{\epsfig{file=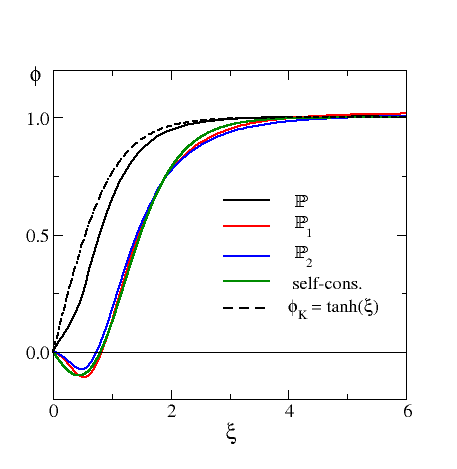,width=7cm,height=4cm}\hspace{1cm}
\epsfig{file=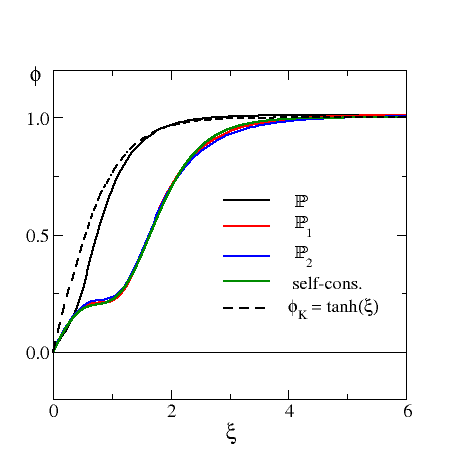,width=7cm,height=4cm}}
\caption{\label{fig1}Profiles for $M=\lambda=2$ and $g=4$. Left panel: $B_1$ configuration,
right panel: $A_1$. The full black line (labeled $\mathbb{P}$) arises from minimizing the total
energy, Eq.\@ (\ref{eq:Etot}). The other lines ($\mathbb{P}_1$, $\mathbb{P}_2$, self-cons.)
are configurations obtained from minimizing the reduced energy functional Eq.(\ref{eq:Ered}).
The legends refer to the notation of tables \ref{tab1} and \ref{tab2}. For completeness we also 
show the kink without any fermion interaction (dashed lines).}
\end{figure}

In table \ref{tab2} we present the results from minimizing $E_{\rm tot}$ using same model parameters 
as in table~\ref{tab1}.
\begin{table}[tb]
\begin{tabular}{c|c|c|c|c}
 &~~~$E_{\rm cl}$~~~&~~~$E_{\rm val}$~~~&~~~$E_{\rm VPE}$~~~&$~~~E_{\rm tot}$~~~\\
\hline
$B_1$ & 1.475 & 2.105 & 1.182 & 4.761 \\
& \multicolumn{4}{|c}{$\mathbb{P}=\left\{-0.472,0.181,1.145,-0.544,0.438\right\}$}\\
\hline\hline
$A_1$ & 1.404 & 3.103 & 1.119 & 5.625 \\
& \multicolumn{4}{|c}{$\mathbb{P}=\left\{-0.264,0.132,0.681,-0.282,0.097\right\}$}
\end{tabular}
\caption{\label{tab2}Results from minimizing the total energy Eq.~(\ref{eq:Etot}) using the 
variational {\it ansatz}, Eq.~(\ref{eq:var1}). Model parameters are as in table \ref{tab1}:
$g=4.0$ and $M=\lambda=2$.}
\end{table}
Obviously there are significant changes when including the VPE into the minimization program. This is not 
unexpected, as we previously found that the VPE is approximately as large as the valence energy. Compared with 
the data in table~\ref{tab1} the total energy decreases by about 10\% by lowering the classical part more strongly 
than increasing the fermion contribution. Yet, we still have $E_{\rm tot}>\frac{Mg}{2}$, which is the mass of 
a free fermion in this case. Hence the soliton configuration is not kinematically stable against
a decay into a free fermion. For the model parameters in tables \ref{tab1} and~\ref{tab2} the 
classical energy of the kink without back-reaction from the fermions is $\frac{4}{3}$ and the 
results from those tables suggests that the VPE strongly mitigates the back-reaction obtained earlier from 
only the valence levels. This also shows up in the graphical representation of the scalar profiles
in figure \ref{fig1}. The profiles for which we found the minimal $E_{\rm tot}$ are labeled $\mathbb{P}$.
They do not differ from the kink $\phi_{\rm K}$ substantially.

We will now elaborate on our numerical analysis of scanning the $g$-$\alpha$ space in more detail. With the 
VPE included even the $A_0$ configuration has a non-zero fermion energy\footnote{The $A_0$ bound state by 
itself is a zero mode with either $u$ or $v$ vanishing so that $\overline{\psi}\psi\big|_{A_0}=0$ and thus 
does not couple to the scalar field.} and thus a deviation $\phi\ne\phi_{\rm K}$ is expected. We will discuss 
this case first although a very similar setting has already been considered in Ref.~\cite{Farhi:2000ws}.
\begin{table}[b]
\begin{tabular}{c||c|c|c||c}
$\alpha$ &~~~~~$E_{\rm cl}$~~~~~~&~~~~$E_{\rm VPE}$~~~~
&~~$E_{\rm tot}$~~&~~$E^{\rm (K)}_{\rm tot}$~~\\
\hline
1.0 & 1.385 & 0.817 & 2.202 & 2.277\\
& \multicolumn{3}{|c||}{$\mathbb{P}=\left\{1.053,1.219,1.104,1.105,0.946\right\}$}\\
\hline
1.2 & 1.648 & 0.829 & 2.477 & 2.543\\
& \multicolumn{3}{|c||}{$\mathbb{P}=\left\{1.053,1.191,1.112,1.094,0.976\right\}$}\\
\hline
1.5 & 2.043 & 0.843 & 2.886 & 2.943\\
& \multicolumn{3}{|c||}{$\mathbb{P}=\left\{1.105,1.159,1.138,1.130,1.032\right\}$}\\
\hline
2.0 & 2.703 & 0.861 & 3.564 & 3.610\\
& \multicolumn{3}{|c||}{$\mathbb{P}=\left\{1.104,1.139,1.120,1.122,1.051\right\}$}\\
\end{tabular}\hspace{1.5cm}
\begin{tabular}{c||c|c|c||c}
$\alpha$ &~~~~~$E_{\rm cl}$~~~~~~&~~~~$E_{\rm VPE}$~~~~
&~~$E_{\rm tot}$~~&~~$E^{\rm (K)}_{\rm tot}$~~\\
\hline
1.0 & 1.437 & 1.079 & 2.516 & 2.693\\
& \multicolumn{3}{|c||}{$\mathbb{P}=\left\{1.093,1.243,1.208,1.157,0.945\right\}$}\\
\hline
1.2 & 1.699 & 1.103 & 2.802 & 2.960\\
& \multicolumn{3}{|c||}{$\mathbb{P}=\left\{1.098,1.243,1.169,1.158,0.951\right\}$}\\
\hline
1.5 & 2.092 & 1.131 & 3.222 & 3.360\\
& \multicolumn{3}{|c||}{$\mathbb{P}=\left\{1.110,1.221,1.162,1.149,0.980\right\}$}\\
\hline
2.0 & 2.748 & 1.167 & 3.914 & 4.026\\
& \multicolumn{3}{|c||}{$\mathbb{P}=\left\{1.106,1.202,1.125,1.146,0.997\right\}$}
\end{tabular}
\caption{\label{tab:A0mode}
Variational approach for the $A_0$ configuration. Left panel: $g=4$, right panel: $g=5$. 
The valence contribution is not listed as $E_{\rm val}\equiv0$ for this configuration. For 
comparison we also list the total energy for the kink profile $\phi_{\rm K}(\xi)=\tanh(\xi)$.} 
\end{table}
For the $A_0$ configuration the kink, $\phi_{\rm K}(\xi)=\tanh(\xi)$ has 
$E^{\rm (K)}_{\rm tot}=\frac{M}{2}\left[\frac{4}{3}\alpha+\epsilon_{\rm VPE}\right]$, where 
$\epsilon_{\rm VPE}=0.943$ and $\epsilon_{\rm VPE}=1.360$ for $g=4$ and $g=5$, respectively.
For the cases shown in table \ref{tab:A0mode} the variational minimum is only slightly less than 
the total energy of the kink. Again, this corroborates the assertion that the inclusion of the 
fermion fields only leads to a moderate back-reaction. We also see that $E_{\rm tot}< \frac{Mg}{2}$
for the $A_0$ case which is thus kinematically stable against a decay into a free fermion. 
However, since the zero mode does not have definite fermion charge, this configuration cannot 
be asserted a particle number, in contrast to a free fermion.

Next we turn to the $B_1$ configuration with the numerical results displayed in table \ref{tab:B1mode}.
\begin{table}[b]
\begin{tabular}{c||c|c|c|c||c}
$\alpha$ &~~~$E_{\rm cl}$~~~&~~~$E_{\rm val}$~~~&~~~$E_{\rm VPE}$~~~
&~~~$E_{\rm tot}$~~~&~~$E^{\rm (K)}_{\rm tot}$~~\\
\hline
1.0 & 1.475 & 2.105 & 1.182 & 4.761 & 4.922 \\
& \multicolumn{4}{|c||}{$\mathbb{P}=\left\{-0.472,0.181,1.145,-0.544,0.438\right\}$}\\
\hline
1.2 & 1.723 & 2.186 & 1.143 & 5.052 & 5.189 \\
& \multicolumn{4}{|c||}{$\mathbb{P}=\left\{-0.472,0.212,1.140,-0.533,0.439\right\}$}\\
\hline
1.5 & 2.104 & 2.270 & 1.103 & 5.477 & 5.589 \\
& \multicolumn{4}{|c||}{$\mathbb{P}=\left\{-0.474,0.243,1.136,-0.522,0.436\right\}$}\\
\hline
2.0 & 2.748 & 2.358 & 1.064 & 6.170 & 6.256 \\
& \multicolumn{4}{|c||}{$\mathbb{P}=\left\{-0.466,0.280,1.112,-0.504,0.434\right\}$}\\
\end{tabular}\hspace{1cm}
\begin{tabular}{c||c|c|c|c||c}
$\alpha$ &~~~$E_{\rm cl}$~~~&~~~$E_{\rm val}$~~~&~~~$E_{\rm VPE}$~~~
&~~~$E_{\rm tot}$~~~&~~$E^{\rm (K)}_{\rm tot}$~~\\
\hline
1.0 & 1.485 & 2.390 & 1.647 & 5.522 & 5.693\\
& \multicolumn{4}{|c||}{$\mathbb{P}=\left\{-0.479,0.132,1.224,-0.544,0.321\right\}$}\\
\hline
1.2 & 1.730 & 2.481 & 1.603 & 5.814 & 5.960\\
& \multicolumn{4}{|c||}{$\mathbb{P}=\left\{-0.474,0.159,1.217,-0.533,0.335\right\}$}\\
\hline
1.5 & 2.107 & 2.576 & 1.559 & 6.241 & 6.360\\
& \multicolumn{4}{|c||}{$\mathbb{P}=\left\{-0.473,0.194,1.203,-0.525,0.358\right\}$}\\
\hline
2.0 & 2.749 & 2.675 & 1.512 & 6.936 & 7.026\\
& \multicolumn{4}{|c||}{$\mathbb{P}=\left\{-0.471,0.237,1.174,-0.516,0.380\right\}$}
\end{tabular}
\caption{\label{tab:B1mode}
Variational approach for the $B_1$ configuration. Left panel: $g=4$, right panel: $g=5$.
The entry $g=4$ and $\alpha=1$ is that of table \ref{tab2}.}
\end{table}
In this case the valence energy is substantial. However, since a strongly distorted kink
comes with large $E_{\rm cl}$ and $E_{\rm VPE}$, {\it cf.\@} table \ref{tab1}, minimizing the 
valence energy does not lower the total energy. Rather configurations that minimize the total energy
have significantly larger valence energies than the solutions constructed in 
Refs.\@ \cite{Klimashonok:2019iya,Perapechka:2019vqv,Saadatmand:2022htx}. The resulting 
configuration is similar to the kink as the variational parameters approximately obey $a\approx d$ 
and $b\approx e$. We find that $c$ is slightly larger than unity which would indicate that profile 
is a squeezed kink. However, the remaining differences between $a$ and $d$ as well as between $b$ 
and $e$ lead to a stretched kink. This is seen in figure~\ref{fig1} in which we show minimizing 
profiles for various channels. As we increase $\alpha$, the classical contribution
to the energy becomes more dominant and the minimizing configuration should get even
closer to the kink. Indeed the numerical simulations confirm that as we find the 
variational parameters 
$\mathbb{P}=\left\{-0.401,0.253,1.070,-0.416,0.342\right\}$ and
$\mathbb{P}=\left\{-0.396,0.274,1.041,-0.408,0.339\right\}$ for $\alpha=3.0$ and 
$\alpha=4.0$, respectively; with $g=4.0$ in both cases. The respective energies
are $E_{\rm tot}=7.532$ and $E_{\rm tot}=8.879$. The discrepancies to the energy of
the kink configuration ($7.589$ and $8.922$) are marginal and decrease as 
$\alpha$ increases.

As a final example we consider the $A_1$ configuration when the valence fermion dwells in 
the third bound state, the second one with even parity. The numerical results are displayed 
in table \ref{tab:A1mode}.
\begin{table}[t]
\begin{tabular}{c||c|c|c|c||c}
$\alpha$ &~~~$E_{\rm cl}$~~~&~~~$E_{\rm val}$~~~&~~~$E_{\rm VPE}$~~~
&~~~$E_{\rm tot}$~~~&~~$E^{\rm (K)}_{\rm tot}$~~\\
\hline
1.0 & 1.404 & 3.103 & 1.119 & 5.625 & 5.741 \\
& \multicolumn{4}{|c||}{$\mathbb{P}=\left\{-0.264,0.132,0.681,-0.282,0.097\right\}$}\\
\hline
1.2 & 1.669 & 3.136 & 1.101 & 5.905 & 6.007 \\
& \multicolumn{4}{|c||}{$\mathbb{P}=\left\{-0.262,0.133,0.702,-0.281,0.102\right\}$}\\
\hline
1.5 & 2.065 & 3.176 & 1.079 & 6.319 & 6.407 \\
& \multicolumn{4}{|c||}{$\mathbb{P}=\left\{-0.260,0.139,0.728,-0.277,0.109\right\}$}\\
\hline
2.0 & 2.724 & 3.228 & 1.051 & 7.003 & 7.074 \\
& \multicolumn{4}{|c||}{$\mathbb{P}=\left\{-0.258,0.141,0.769,-0.271,0.116\right\}$}
\end{tabular}\hspace{1cm}
\begin{tabular}{c||c|c|c|c||c}
$\alpha$ &~~~$E_{\rm cl}$~~~&~~~$E_{\rm val}$~~~&~~~$E_{\rm VPE}$~~~
&~~~$E_{\rm tot}$~~~&~~$E^{\rm (K)}_{\rm tot}$~~\\
\hline
1.0 & 1.381 & 3.663 & 1.562 & 6.606 & 6.693 \\
& \multicolumn{4}{|c||}{$\mathbb{P}=\left\{-0.248,0.081,0.749,-0.263,0.063\right\}$}\\
\hline
1.2 & 1.645 & 3.697 & 1.541 & 6.882 & 6.960 \\
& \multicolumn{4}{|c||}{$\mathbb{P}=\left\{-0.255,0.107,0.766,-0.263,0.083\right\}$}\\
\hline
1.5 & 2.044 & 3.725 & 1.523 & 7.292 & 7.360 \\
& \multicolumn{4}{|c||}{$\mathbb{P}=\left\{-0.252,0.105,0.781,-0.263,0.085\right\}$}\\
\hline
2.0 & 2.706 & 3.768 & 1.496 & 7.970 & 8.026 \\
& \multicolumn{4}{|c||}{$\mathbb{P}=\left\{-0.245,0.095,0.814,-0.255,0.081\right\}$}
\end{tabular}
\caption{\label{tab:A1mode}
Variational approach for the $A_1$ configuration. Left panel: $g=4$, right panel: $g=5$.
The entry $g=4$ and $\alpha=1$ is that of table \ref{tab2}.}
\end{table}
The scenario is pretty much the same as for the previous configurations: The inclusion
of the fermion VPE moves the distorted kink back to a slightly stretched kink. We see 
that even for $\alpha=1$ the variational solution is not very different from the kink.
This implies that the total energy is dominated by the classical part which is another
indication that there are sizable cancellations between the gain from binding a valence 
fermion and the fermion VPE. All our variational searches yield $a>d$ so that the profile 
approaches the vacuum expectation value from above as $\xi\,\to\,\infty$.

\begin{figure}[tb]
\centerline{\epsfig{file=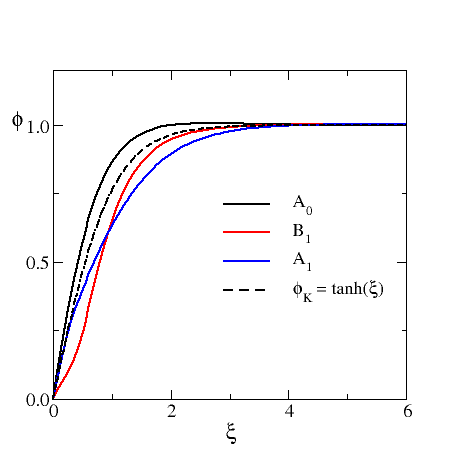,width=10cm,height=4cm}}
\caption{\label{fig:solutions}
Variational solutions of the $A_0$, $B_1$ and $A_1$ configurations for $\alpha=1.0$ and $g=4.0$.
For comparison the kink profile is also shown.}
\end{figure}

For a certain parameter set we display the profiles that minimize $E_{\rm tot}$ in the
various channels in Fig.\@ \ref{fig:solutions}.
When compared to the kink, the solution of the $A_0$ configuration is squeezed, while the others 
are stretched. The reason is that squeezing the kink slightly decreases the VPE. To considerably gain
energy from binding a non-zero mode, the Yukawa interaction must be strongly attractive which
requires an extended kink profile. That is what we observe for the $B_1$ and $A_1$ configurations.

\begin{figure}[h]
\centerline{\epsfig{file=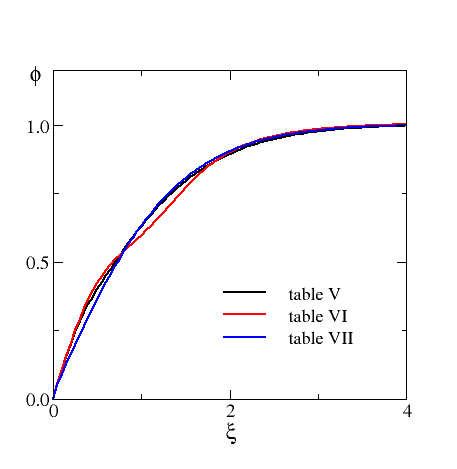,width=7cm,height=4cm}\hspace{1cm}
\epsfig{file=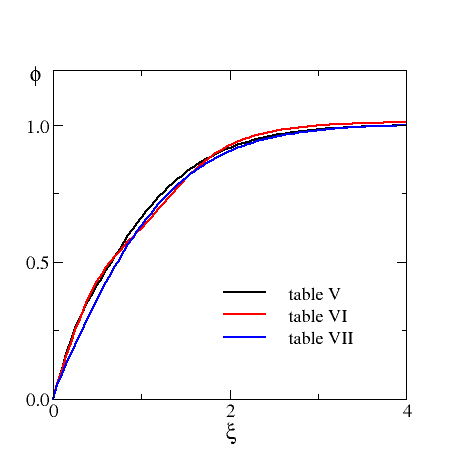,width=7cm,height=4cm}}
\caption{\label{fig2}Comparison of the different solutions for the $A_1$ scenario according to the
respective tables. We always have $\alpha=1$ as in table \ref{tab1} and figure \ref{fig1}. 
Left panel $g=4$, right panel $g=5$.}
\end{figure}

We should, however, mention that we find various local minima for the total energy in 
the space of the chosen variational parameters. Their total energies may be slightly smaller
or larger than those reported above; but they turn out to be stationary under the steepest 
descent algorithm. This is the case for all configurations
but we will discuss this only for the $A_1$ configuration in more detail. The data for some of
the alternative solutions are listed in tables \ref{tab:A1mode_alt} and \ref{tab:A1mode_alt2}.
Looking at the corresponding variational parameters suggests that these solutions
would be quite different. In particular those from table \ref{tab:A1mode_alt}
exhibit significant deviations from the kink relation $b\approx e$.
However, when plotting them, {\it cf.\@} Fig.\@ \ref{fig2}, we observe that the
profiles are approximately identical and differences get smaller as the Yukawa coupling 
increases. We conjecture that these solutions are pretty close to the actual solution, 
but the particular variational ansatz is not able of capturing it exactly. Furthermore the 
minimum may be very shallow. The ansatz most likely does not allow for a continuous 
transition between the solutions on a path that can be constructed from a steepest descent
procedure. In any event, we have sufficient evidence to state that the actual solution 
will be similar to the kink but quite distinct from the solutions of 
Refs.\@ \cite{Klimashonok:2019iya,Perapechka:2019vqv,Saadatmand:2022htx}.
The latter could emerge for model parameters for which the classical and/or Dirac sea contributions, 
which both prefer the standard kink profile, are suppressed compared to the valence part. 
Eq.\@ (\ref{eq:Etot}) reveals that such parameters are governed by small $\alpha$ 
for which the present semi-classical expansion is not reliable.
\begin{table}[t]
\begin{tabular}{c||c|c|c|c}
$\alpha$ &~~~$E_{\rm cl}$~~~&~~~$E_{\rm val}$~~~&~~~$E_{\rm VPE}$~~~&$~~~E_{\rm tot}$~~~\\
\hline
1.0 & 1.443 & 3.035 & 1.141 & 5.618 \\
& \multicolumn{4}{|c}{$\mathbb{P}=\left\{-1.963,1.521,1.345,-2.060,1.875\right\}$}\\
\hline
1.2 & 1.696 & 3.092 & 1.113 & 5.900 \\
& \multicolumn{4}{|c}{$\mathbb{P}=\left\{-1.917,1.530,1.268,-2.002,1.848\right\}$}\\
\hline
1.5 & 2.073 & 3.156 & 1.088 & 6.317 \\
& \multicolumn{4}{|c}{$\mathbb{P}=\left\{-1.832,1.551,1.141,-1.896,1.807\right\}$}\\
\hline
2.0 & 2.733 & 3.220 & 1.051 & 7.004 \\
& \multicolumn{4}{|c}{$\mathbb{P}=\left\{-1.846,1.580,1.154,-1.905,1.807\right\}$}\\
\end{tabular}\hspace{2cm}
\begin{tabular}{c||c|c|c|c}
$\alpha$ &~~~$E_{\rm cl}$~~~&~~~$E_{\rm val}$~~~&~~~$E_{\rm VPE}$~~~&$~~~E_{\rm tot}$~~~\\
\hline
1.0 & 1.419 & 3.507 & 1.644 & 6.571 \\
& \multicolumn{4}{|c}{$\mathbb{P}=\left\{-1.664,0.995,1.245,-1.727,1.120\right\}$}\\
\hline
1.2 & 1.660 & 3.634 & 1.586 & 6.879 \\
& \multicolumn{4}{|c}{$\mathbb{P}=\left\{-1.833,1.457,1.165,-1.918,1.719\right\}$}\\
\hline
1.5 & 2.052 & 3.692 & 1.549 & 7.293 \\
& \multicolumn{4}{|c}{$\mathbb{P}=\left\{-1.824,1.519,1.116,-1.898,1.744\right\}$}\\
\hline
2.0 & 2.710 & 3.746 & 1.514 & 7.971 \\
& \multicolumn{4}{|c}{$\mathbb{P}=\left\{-1.748,1.490,1.107,-1.804,1.666\right\}$}
\end{tabular}
\caption{\label{tab:A1mode_alt} Alternative solutions for the $A_1$ configuration: 
Left panel $g=4$, right panel $g=5$.}
\end{table}
\begin{table}[t]
\begin{tabular}{c||c|c|c|c}
$\alpha$ &~~~$E_{\rm cl}$~~~&~~~$E_{\rm val}$~~~&~~~$E_{\rm VPE}$~~~&$~~~E_{\rm tot}$~~~\\
\hline
1.0 & 1.394 & 3.109 & 1.138 & 5.642 \\
& \multicolumn{4}{|c}{$\mathbb{P}=\left\{2.600,0.170,0.717,2.581,0.088\right\}$}\\
\hline
1.2 & 1.660 & 3.144 & 1.115 & 5.918 \\
& \multicolumn{4}{|c}{$\mathbb{P}=\left\{2.441,0.207,0.734,2.428,0.117\right\}$}\\
\hline
1.5 & 2.057 & 3.185 & 1.090 & 6.332 \\
& \multicolumn{4}{|c}{$\mathbb{P}=\left\{2.590,0.182,0.766,2.579,0.110\right\}$}\\
\hline
2.0 & 2.716 & 3.239 & 1.058 & 7.013 \\
& \multicolumn{4}{|c}{$\mathbb{P}=\left\{2.558,0.188,0.805,2.575,0.126\right\}$}
\end{tabular}\hspace{2cm}
\begin{tabular}{c||c|c|c|c} 
$\alpha$ &~~~$E_{\rm cl}$~~~&~~~$E_{\rm val}$~~~&~~~$E_{\rm VPE}$~~~&$~~~E_{\rm tot}$~~~\\
\hline
1.0 & 1.370 & 3.667 & 1.578 & 6.621\\
& \multicolumn{4}{|c}{$\mathbb{P}=\left\{2.609,0.143,0.770,2.578,0.074\right\}$}\\
\hline
1.2 & 1.637 & 3.694 & 1.561 & 6.892\\
& \multicolumn{4}{|c}{$\mathbb{P}=\left\{2.449,0.193,0.780,2.428,0.108\right\}$}\\
\hline
1.5 & 2.036 & 3.728 & 1.536 & 7.300\\
& \multicolumn{4}{|c}{$\mathbb{P}=\left\{2.442,0.193,0.801,2.425,0.117\right\}$}\\
\hline
2.0 & 2.700 & 3.771 & 1.507 &7.978\\
& \multicolumn{4}{|c}{$\mathbb{P}=\left\{2.592,0.155,0.834,2.577,0.103\right\}$}
\end{tabular}
\caption{\label{tab:A1mode_alt2} Second set of alternative solutions for the $A_1$ configuration: 
left panel $g=4$ right panel $g=5$.}
\end{table}

Without the fermion coupling, the boson contribution to the VPE is $\frac{1}{2\sqrt{3}}-\frac{3}{\pi}\approx-0.666$
for $M=2$ \cite{Dashen:1974cj,Ra82,Graham:2009zz}. With that coupling included the boson VPE is difficult to estimate 
because there is a linear term in the harmonic approximation for the fluctuations when the profile is not 
a solution to the classical kink equation. Furthermore imaginary frequency eigenvalues emerge for the boson 
fluctuations~\cite{Graham:1998kz}. Yet, we assume that the above number is a useful estimate because the variational 
profiles are quite similar to the kink. As conjectured earlier, the numerical simulations verify that the fermion 
VPE becomes significantly larger when the Yukawa coupling increases. In subsection \ref{sec:FCon} we especially argued 
that for large enough values of the Yukawa coupling constant the fermion VPE should dominate the boson counterpart.
And indeed we always find that $E_{\rm VPE}>\frac{M}{2}$ when $g\ge5$.

\section{Conclusions}
\label{sec:Con}

In this project we have considered a model with fermions coupled to the kink in one space and one time dimension 
and investigated the fermion back-reaction on the kink profile. In this approach a valence fermion appears as an
explicitly occupied bound state level and the back-reaction may be significant when this is not the ground state
level. And indeed earlier studies, that only considered the coupling of the kink to a single fermion valence level, 
found considerable back-reactions. However, it is important to not only consider the energy of the occupied valence
level but also to add the energy of the Dirac sea to the functional that determines the back-reaction. The main 
argument for its inclusion is the consistency of the semi-classical expansion. This contribution is obtained as 
the renormalized sum of the changes of the one-particle fermion energies. These changes emerge because the
Yukawa coupling between the kink and the fermions polarizes the fermion vacuum. They concern the discrete bound 
states as well as the continuous scattering states. The boson contribution to the vacuum polarization energy has not 
been included in the energy functional. Rather we have argued that, for large values of the Yukawa coupling constant, 
the fermion energies would dominate the quantum corrections. We have verified that conjecture {\it a posteriori}.

We have utilized a variational approach to find (an upper bound) to the minimal total energy. We have only 
considered a single variational parameterization of the kink profile with five parameters to lessen the 
numerical efforts. Certainly this leaves space for improvement. Nevertheless, we consider it sufficient to 
show that the Dirac sea contribution brings back the strongly distorted kink profiles from the self-consistent 
treatment for the case that omits the Dirac sea, to a slightly squeezed or stretched kink profile. The similarity 
to the kink is particularly pronounced for the $A_0$ configuration where we do not have a valence quark 
contribution to the energy. For that reason, a slightly squeezed kink profile is energetically favorable for 
the $A_0$ configuration. In all other cases, some energy is gained from binding the valence level by stretching 
the kink because it makes the Yukawa interaction more attractive. Since this deformation increases both the 
classical energy and the vacuum polarization energy, the stretch can only be modest.

The dominant effect for mitigating the kink distortion is the balance between the fermion bound state and vacuum 
energies so that the profile function is governed by minimizing the classical energy. This balance could well 
be a consequence of Levinson's theorem \cite{Levinson:1949,Barton:1984py}, which 
relates the number of bound states to the density of continuum states. Essentially it states that any bound 
state must have emerged from the continuum so that there is an exact balance between bound and scattering states
when the background potential changes. The fermion energy has two contributions, the valence part which arises 
from the bound state energies and the vacuum polarization energy, which results from the scattering data. For
the energies the balance is not exact because the particle number (density) carries factors of the single 
particle energies. Nevertheless the theorem strongly suggests that the energy gain from one comes with an energy 
loss of the other.

Though kink type solutions emerge in many field theories, systems with explicit scalar and fermion fields are
most prominent in variants of the standard model of particle physics. When embedded in three space dimensions, 
kinks represent domain walls in those models \cite{Vilenkin:2000jqa}. In these models the fermions' back-reaction
on the kink may eventually exhibit effects like those that we have explored here, in particular when the fermions 
are bound by the domain wall and dwell in excited levels.

We have considered cases in which only a single fermion occupation number was non-zero. The valence energy 
should become more important when this is true for several occupation numbers. Then we expect a stronger
distortion of the kink. This should be subject of future projects.

\acknowledgments
H.\@ W.\@ is supported in part by the National Research Foundation of South Africa (NRF) by grant~150672.

\end{document}